\documentclass[12pt]{article}

\catcode`\@=11

\global\arraycolsep=2pt
\usepackage{amsbsy,amssymb,latexsym,amsfonts,amsmath}
\usepackage{graphicx,color}

\begin{document}

\begin{flushright}
YITP-24-51
\end{flushright}

\vspace*{0.7cm}

\begin{center}
{ \Large  Who told you magnetization is a vector in $4-\epsilon$ dimensions?}
\vspace*{1.5cm}\\
{Yu Nakayama}
\end{center}
\vspace*{1.0cm}
\begin{center}

Yukawa Institute for Theoretical Physics,
Kyoto University, Kitashirakawa Oiwakecho, Sakyo-ku, Kyoto 606-8502, Japan

\vspace{3.8cm}
\end{center}

\begin{abstract}
But if you treat it as a two-form, you get three nontrivial renormalization group fixed points! Which becomes the Heisenberg fixed point in three dimensions? Motivated by this question, we study the conformal bootstrap constraint in the $O(d)$ anti-symmetric matrix model in $d$ dimensions, varying $d$ as a continuous parameter. Besides the one that is naturally connected to the Heisenberg fixed point in three dimensions, we find ``evanescent" kinks whose origin is yet to be identified. We also bootstrap $O(4), O(5), O(6)$  anti-symmetric matrix model in $d=3$, aiming at physical applications.
\end{abstract}

\thispagestyle{empty} 

\setcounter{page}{0}

\newpage

\section{Introduction}
Since we live in $1+3$ dimensional space-time, we are used to the idea that electric field and magnetic field are both three-vectors, so much so that we sometimes treat the magnetic field in other dimensions as if it were a vector. 
In generic dimensions, the magnetic field and therefore a magnetization is rather a two-form. Treating it as a vector must be regarded as a similar sin to the idea that the rotation can be described by a rotation vector in higher dimensions.

When we learn the $\epsilon$ expansions \cite{Wilson:1973jj} to estimate the critical exponents of the Heisenberg model in three dimensions, we typically consider the $O(3)$ vector model in $4-\epsilon$ dimensions. Here, we treat the magnetization as an internal $O(3)$ vector rather than a space vector. The origin of $O(3)$ ``rotational" symmetry in $4-\epsilon$ dimension, however, is not immediately obvious. Was it more appropriate to study the $O(4-\epsilon)$ two-form (or anti-symmetric matrix) model in $4-\epsilon$ dimensions?

Of course, we can argue that the $\epsilon$ expansion is just a mathematical trick and we should not speak of its physical meaning. There are infinitely different ways to approach the Heisenberg fixed point by changing the dimensionality, and in a certain sense, there is no ``physically more reasonable" model.
Still, it seems an interesting question to ask what happens if we study the $O(4-\epsilon)$ two-form (or anti-symmetric matrix) model in $4-\epsilon$ dimensions rather than the $O(3)$ vector model? It may give an intrinsic ambiguity in the predictions of the $\epsilon$ expansions by taking the $\epsilon \to 1$ limit. A related question has been addressed when we study (space-time) ``spinors" in $4-\epsilon$ dimensions. The spinor representation in non-integer dimensions is not well-defined and many different prescriptions exist.
The role of so-called evanescent operators there has not been completely understood (see e.g. \cite{Ji:2018emi}). See also \cite{Cao:2023psi} for the evanescent operators with $O(N)$ symmetry, which are relevant to our study.

We approach this question by using the non-perturbative conformal bootstrap. While bootstrapping conformal field theories with physical symmetry in physical dimensions (e.g. $O(3)$ symmetric Heisenberg model in three dimensions) has achieved tremendous success, in determining critical exponents and operator product expansions (OPE) coefficients with unparalleled accuracy \cite{ElShowk:2012ht}\cite{El-Showk:2014dwa}\cite{Kos:2014bka}\cite{Simmons-Duffin:2016wlq}\cite{Kos:2016ysd}, interesting insights have been obtained by studying conformal bootstrap with unphysical symmetry, such as $O(N)$ with non-integer $N$, in non-integer dimensions \cite{El-Showk:2013nia}\cite{Shimada:2015gda}\cite{Cappelli:2018vir}\cite{Sirois:2022vth}\cite{Henriksson:2022gpa}\cite{Bonanno:2022ztf}\cite{Hogervorst:2015akt}\cite{Golden:2014oqa}. In a similar manner, keeping track of the candidate of the conformal fixed point predicted by the conformal bootstrap while changing the dimensionality $d$, we will gain a non-perturbative understanding of what happens to the $O(4-\epsilon)$ two-form model in $4-\epsilon$ dimensions in the limit of $\epsilon \to 3$.

Since there are hundreds of papers working on conformal bootstrap, we only quote some relevant ones whose main target is the Landau-Ginzburg model in various dimensions \cite{Nakayama:2014lva}\cite{Nakayama:2014yia}\cite{Nakayama:2014sba}\cite{Bae:2014hia}
\cite{Chester:2014gqa}\cite{Kos:2015mba}\cite{Nakayama:2016jhq}\cite{Li:2016wdp}\cite{Nakayama:2017vdd}\cite{Rong:2017cow}\cite{Stergiou:2018gjj}\cite{Kousvos:2018rhl}\cite{Go:2019lke}\cite{Stergiou:2019dcv}\cite{Kousvos:2019hgc}\cite{Chester:2019ifh}\cite{Henriksson:2020fqi}\cite{Dowens:2020cua}\cite{Chester:2020iyt}\cite{Reehorst:2020phk}\cite{Manenti:2021elk}\cite{Reehorst:2021ykw}\cite{Kousvos:2021rar}\cite{Kousvos:2022ewl}\cite{Liu:2023elz}. See e.g. reviews \cite{Poland:2018epd}\cite{Poland:2022qrs}\cite{Rychkov:2023wsd} and lecture notes \cite{Simmons-Duffin:2016gjk}\cite{Chester:2019wfx} for a more complete list. We also give some new bounds on conformal dimensions of operators in integer dimensions with physical symmetry.

The organization of the paper is as follows. In section 2, we introduce the $O(N)$ anti-symmetric matrix model within the $\epsilon$ expansions. In section 3, we set up the conformal bootstrap equations to study the constraint on the conformal dimensions of various spin-zero operators. We also provide results of numerical conformal bootstrap in related models for comparison. In section 4, we present further discussions and conclude.

\section{$O(N)$ anti-symmetric matrix model}
Consider the $O(N)$ anti-symmetric matrix model, where a real anti-symmetric $N$-by-$N$ matrix $\phi$ is a dynamical variable. We assume the $O(N)$ global symmetry and $\phi$ transforms as an adjoint representation of the $O(N)$ symmetry. The action is 
\begin{align}
S = \int d^dx -\frac{1}{2}\mathrm{Tr} (\partial_\mu \phi \partial^\mu \phi ) - \frac{m^2}{2} \mathrm{Tr}\phi^2 + \frac{\lambda_1}{4!} (\mathrm{Tr} (\phi^2))^2 + \frac{\lambda_2}{4!} \mathrm{Tr} (\phi^4) \ . 
\end{align}
In the following, we will always assume that we fine-tune $m^2$ so that it is located at its critical value. Within the first-order perturbation theory in $4-\epsilon$ dimensions, the renormalization group beta functions are given by
\begin{align}
\beta_{\lambda_1} &= -\epsilon \lambda_1 + \frac{\lambda_1^2}{6}(N^2-N+16) + \frac{\lambda_1 \lambda_2}{3}(2N-1) + \frac{\lambda_2^2}{2} + O(\lambda^3) \cr
\beta_{\lambda_2} &= -\epsilon \lambda_2 + 4 \lambda_1 \lambda_2 +  \frac{\lambda_2^2 }{6}(2N-1) + O(\lambda^3)  \  . 
\end{align}

\begin{figure}[htbp]
	\begin{center}
		\includegraphics[width=7.8cm,clip]
  {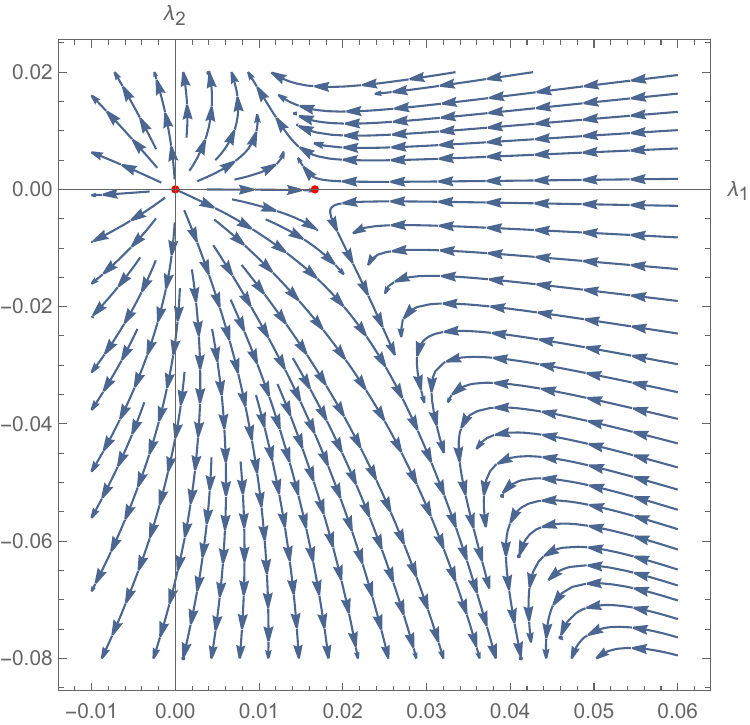}
  \includegraphics[width=7.8cm,clip]
  {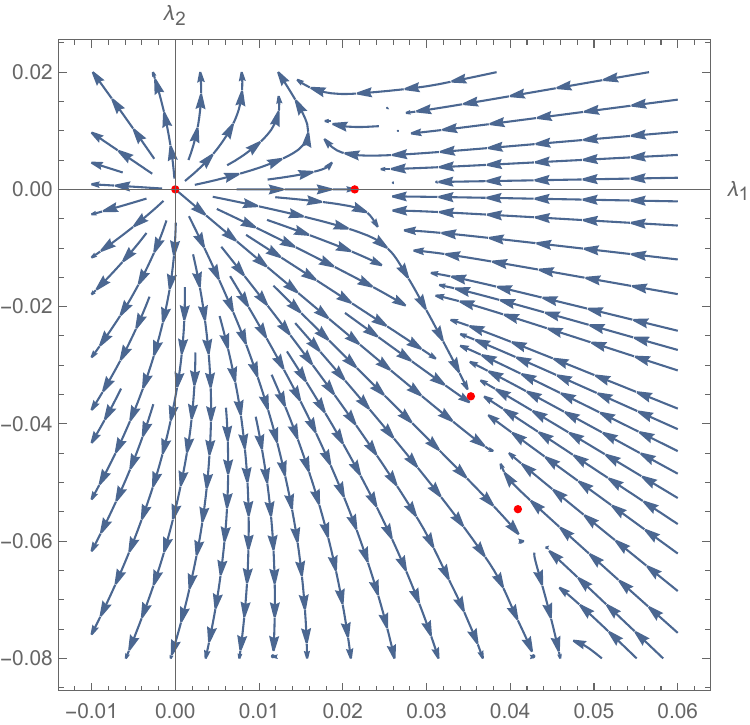}
    \includegraphics[width=7.8cm,clip]
  {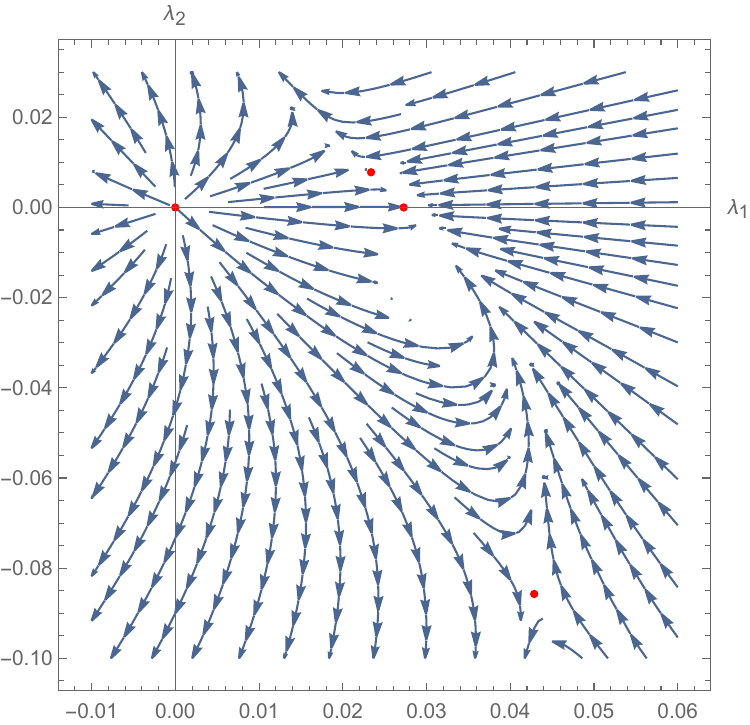}
	\end{center}
	\caption{One-loop renormalization group flow in $O(5)$ (top), $O(4)$ (middle), and $O(3)$ (bottom) anti-symmetric matrix model in $d=3.9$. Red dots are real fixed points predicted in the $\epsilon$ expansions.}
	\label{fig:rg1}
\end{figure}

In the small $\epsilon$ limit, there are four renormalization group fixed points $\beta_{\lambda_1}=\beta_{\lambda_2}= 0$. Beside the trivial Gaussian fixed point $(\lambda_1,\lambda_2) = (0,0)$ and the $O(N(N-1)/2)$ symmetric Wilson-Fisher fixed point\footnote{The $O(N)$ anti-symmetric matrix $\phi$ becomes $O(N(N-1)/2)$ vector $\vec{\phi}$. Note $\mathrm{Tr}\phi^2 = -\vec{\phi} \cdot \vec{\phi}$. }$(\lambda_1,\lambda_2) = (\frac{6}{N^2-N+16} \epsilon,0)$ (we call it the vector model fixed point $V$), we have two genuine $O(N)$ symmetric fixed points (we call them the matrix model fixed points $M_{\pm}$) located at
\begin{align} 
\lambda_1 &= \frac{3 \left(-4 N^2+4 N+143  \mp \sqrt{-(2 N-1)^2 \left(8 N^2-8 N-97\right)}\right)}{4 N^4-8 N^3-123 N^2+127 N+1696} \epsilon\cr
\lambda_2 &= \frac{6 \left(4 N^4-8 N^3-75 N^2 +79 N-20 \pm 12 \sqrt{-(2 N-1)^2 \left(8 N^2-8 N-97\right)}\right)}{(2 N-1) \left(4 N^4-8 N^3-123 N^2+127 N+1696\right)} \epsilon
\end{align}
When $N>N_c$, these two fixed points $M_{\pm}$ become complex.\footnote{The formal fixed points with complex coupling constants are known as complex fixed points (see e.g. \cite{Gorbenko:2018ncu} and reference therein). Unlike the mild violation of unitarity for non-integer $d$ and $N$ \cite{Golden:2014oqa}\cite{Hogervorst:2015akt}, the violation of the unitary in complex fixed points is typically large. It is thus expected that the nontrivial features in the conformal bootstrap bound will disappear for $N>N_c$.} In the small $\epsilon$ limit, $N_c =  \frac{1}{4}(2 + 3 \sqrt{22}) \sim 4.01781 $. We show the one-loop renormalization group flow of $O(3)$, $O(4)$ and $O(5)$ anti-symmetric matrix model in fig \ref{fig:rg1}. 

Since in the cases $N=3$ and $N=2$, the anti-symmetric matrix model reduces to an $O(3)$ vector model and the Ising model and there is no genuine anti-symmetric matrix to talk about,  only the case $N=4$ (or possibly $N>4$ as well for larger $\epsilon$) is of direct physical relevance: it is a nontrivial question if the $N=4$ fixed point remains as a real fixed point in three dimensions. Currently, we do not know the precise value of $N_c$ in three dimensions. 

Within the $\epsilon$ expansions, by using the standard perturbation theory, we can compute the anomalous\footnote{``Anomalous" means that the conformal dimensions are measured from the Gaussian fixed point: $\Delta = \frac{2-\epsilon}{2}n + \gamma$, where $n=1$ for $\phi$ and $n=2$ for $S,T,A_4$ and $B_4$.} dimensions of (composite) operators
\begin{align}
\gamma_{\phi} &= \frac{1}{288} \left(N^2-N+4\right) \left(4 {\lambda_1}^2+{\lambda_2}^2\right)+\frac{1}{36} (2 N-1) ({\lambda_1} {\lambda_2}) + O(\lambda^3) \cr
\gamma_{S} &=  \frac{\lambda_1}{6} \left(N^2-N+4\right)+\frac{\lambda_2}{6} (2 N-1) + O(\lambda^2) \cr
\gamma_{T} &=  \frac{\lambda_1}{6} \left(4\right)+\frac{\lambda_2}{6} (N-1) + O(\lambda^2) \cr
\gamma_{A_4} &= \frac{\lambda_1}{6} \left(4\right)-\frac{\lambda_2}{6} (2) + O(\lambda^2)  \cr
\gamma_{B_4} &=  \frac{\lambda_1}{6} \left(4\right)+\frac{\lambda_2}{6} (1) + O(\lambda^2)   \ . \label{perturbative}
\end{align}
Here $S$ is $O(N)$ singlet (i.e. $\mathrm{Tr}(\phi^2)$), and $T$ is a traceless symmetric (rank 2) tensor (i.e. $\sum_b \phi_{ab}\phi_{bc} -\text{trace}$). $A_4$ is a totally antisymmetric rank 4 tensor (i.e. $\phi_{[a,b}\phi_{c,d]})$, and $B_4$ is a two-by-two ``box" representation in terms of the Young Tableau (it has the same symmetry as the Weyl tensor in general relativity). See also footnote \ref{representation} below for our group-theoretic notation.

With the spirit of the $\epsilon$ expansion, as a model of magnetisation, we will study $O(d)$ anti-symmetric matrix model in $d$ dimensions. The above $\epsilon$ expansion tells us that we have three nontrivial fixed points; the one with enhanced $O(N(N-1)/2)$ symmetry (vector model fixed point $V$) and the $O(N)$ symmetric genuine matrix model fixed point $M_{\pm}$. While it may seem natural to expect that the fixed point $V$ will become the Heisenberg fixed point when we set $\epsilon = 1$, it is less obvious what will happen to the other two fixed points. It may seem equally reasonable to keep track of the stable fixed point $M_+$ (rather than $V$) because the Heisenberg fixed point must be reached by fine-tuning only one parameter. This is not only an academic question.
The physical prediction may depend on the choice, in particular, if we reshuffle the perturbation series with respect to $\epsilon = 4-d = 4-N$ simultaneously.\footnote{An alternative possibility is they merge into a complex fixed point at a certain finite value of $\epsilon$.} 

Since the order of limit may matter, let us also discuss the $N\to 3$ limit while keeping $\epsilon$ small. Before reaching $N=3$, something interesting happens at $N = N_*= \frac{1}{2}(1+\sqrt{33}) \sim 3.3723$, where one of the matrix model fixed point $M_+$ collides with the vector model fixed point $V$ on the $\lambda_2=0$ axis. Unlike the case at $N=N_c$, they are not annihilated into the complex coupling constant space, but they remain in the real coupling constant space even after the collision. Yet, the stability structure changes: while above $N_*= \frac{1}{2}(1+\sqrt{33}) \sim 3.3723$ the matrix model fixed point $M_+$ was the most stable fixed point, below $N_*$ the vector model fixed point $V$ becomes the most stable fixed point. See Fig \ref{fig:rg1}. If we stick to the idea that we should keep track of the most stable fixed point, we have to change lanes and the road is non-analytic!

In the limit $N\to 3$, all four fixed points remain real and they appear independent in the coupling constant space, but in the physical sense, the two ``nontrivial" fixed points are equivalent to the other two fixed points. The matrix model fixed point $M_+$ at $(\lambda_1,\lambda_2) = (\frac{18}{77}\epsilon,\frac{6}{77}\epsilon)$ is equivalent to the vector model fixed point  $V$ at $(\frac{3}{11}\epsilon,0)$ (with the same $O(3)$ symmetry)  and the matrix model fixed point $M_-$ at $(\lambda_1,\lambda_2) = (\frac{3}{7}\epsilon,-\frac{6}{7}\epsilon)$ is equivalent to the Gaussian fixed point $(0,0)$. The reason is that for the physical operators such as $\phi$, $S$ or $T$, the anomalous dimensions depend only on the combination $2\lambda_1 + \lambda_2$ in the $N=3$ limit, and they take the same values at both points.
In contrast, the evanescent operators such as $A_4$ or $B_4$, which do not exist in $N=3$, do show different anomalous dimensions at each point. We will come back to this when we discuss the results of the numerical conformal bootstrap in later sections.

We would like to note that the appearance of $N_*$ is somehow related to the cubic instability of the $O(N_V)$ vector model fixed point (here $N_V = \frac{N(N-1)}{2}$). The cubic instability of the vector model has been under debate for many years since the birth of the renormalization group method (see e.g. \cite{Cardy:1996xt}). Here, we observe that the instability is triggered by the same operator in the $O(N_V)$ rank-four tensor representation $T_4$  but with different linear combinations (which vanish in the $d \to 3$ limit). The one-loop anomalous dimension of $T_4$ is $\gamma_{T_4} = \frac{12}{N_V + 8} \epsilon$ and becomes relevant if $N_V< 4 + O(\epsilon)$.
The recent results in conformal bootstrap suggest that $N_V^*<3$ (i.e. $N_* <3$ as well) in $d=3$ \cite{Chester:2020iyt}, so it is likely that $M_+$ remains the most stable fixed point in the $N \to 3$ limit in our problem.  

Thus, after careful resummation of the $O(d-\epsilon)$ anti-symmetric model in $4-\epsilon$ dimensions, both the vector model fixed point $V$ and the matrix model fixed point $M_+$ should approach the same Heisenberg fixed point in the $\epsilon \to 1$ limit without a collision. Such a constraint may be useful to estimate the critical exponents of the Heisenberg fixed point more accurately based on the $\epsilon$ expansions.

While we could pursue the higher order (resumed) perturbation theory to strengthen the discussion (see e.g. \cite{Bednyakov:2021ojn} for the direction), we instead seek to approach them by using the non-perturbative conformal bootstrap method. This will be investigated in the next section.

\section{Bootstrapping $O(N)$ anti-symmetric matrix model}

\subsection{Bootstrap equations}
In this paper, we study the bootstrap equation for the single four-point function $\langle \phi \phi \phi \phi \rangle$, where $\phi$ is a spin-zero operator in the anti-symmetric tensor representation of $O(N)$.
 We decompose the OPE of $\phi\times \phi$ into the irreducible representations\footnote{In terms of the Young Tableau, we have $T = [1,1]$ (whose dimension is $d_T = \frac{(N-1)(N+2)}{2}$) , $A= [2]$ ($d_A = \frac{N(N-1)}{2}$), $A_4 = [4]$ ($d_{A_4} = \frac{N(N-1)(N-2)(N-3)}{24}$), $B_4= [2,2]$ ($d_{B_4} = \frac{N(N+1)(N+2)(N-3)}{12}$), and $SA (= Y_{2,1,1}) =  [3,1] $ ($d_{SA} = \frac{N(N+2)(N-1)(N-3)}{8}$), where the number in the bracket denotes the number of column boxes. \label{representation}} to obtain the OPE sum rule for the four-point functions: The so-called conformal bootstrap equations read
\begin{align}
0 =& \sum_{O \in \phi \times \phi} \lambda_O^2 V^{(+)}_{S}+ \sum_{O \in \phi \times \phi} \lambda_O^2 V^{(+)}_{T} +\sum_{O \in \phi \times \phi}  \lambda_O^2 V^{(+)}_{A_4} \cr
 &+  \sum_{O \in \phi \times \phi} \lambda_O^2 V^{(+)}_{B_4} +  \sum_{O \in \phi \times \phi} \lambda_O^2 V^{(-)}_{A} +  \sum_{O \in \phi \times \phi} \lambda_O^2 V^{(-)}_{SA}
\end{align}
where $(\pm)$ denotes the even $(+)$ or odd $(-)$ spin contributions. By using the convention
\begin{align}
F & = v^{\Delta_{\phi}} g_{\Delta_O,l}(u,v) - u^{\Delta_{\phi}} g_{\Delta_O,l}(v,u) \cr
H & = v^{\Delta_{\phi}} g_{\Delta_O,l}(u,v) + u^{\Delta_{\phi}} g_{\Delta_O,l}(v,u)
\end{align}
with the conformal block $g_{\Delta_O,l}$ being normalized as in \cite{Hogervorst:2013kva}, whose explicit expression can be found in \cite{Dolan:2003hv}, each representation contributes to the sum rule as
\begin{align}
V_{S}^{(+)} &= \left( \begin{array}{cc}  \left(1+\frac{2}{N(N-1)}\right)F  \\  \frac{2}{N(N-1)}F  \\ \frac{2}{N(N-1)}F \\ \frac{2}{N(N-1)}F \\ \frac{2}{N(N-1)}H \\  \frac{2}{N(N-1)}H \\
\end{array} \right) \ , \ \
V_{{T}}^{(+)} = \left( \begin{array}{cc}  \frac{N+2}{N} F \\ \left(1 + \frac{N^2-8}{2N(N-2)}  \right)F \\ \frac{N-4}{N(N-2)}F \\ \frac{-2(N+2)}{N(N-2)} F \\ \frac{(N-4)(N+2)}{2N(N-2)}H\\  \frac{-4}{N(N-2)} H \\
\end{array} \right) \ , \ \
V_{A_4}^{(+)} = \left( \begin{array}{cc}  \frac{(N-3)(N-2)}{12} F \\ \frac{3-N}{6} F \\ \frac{1}{6}F \\ \frac{7}{6} F \\ \frac{N-3}{6} H \\  -\frac{1}{6}H \\
\end{array} \right) \ , \cr
V^{(+)}_{B_4} &=  \left( \begin{array}{cc}  \frac{(N-3)(N+1)(N+2)}{6(N-1)} F \\ \frac{(N-4)(N-3)(N+1)}{6(N-2)(N-1)} F \\ \left(1+\frac{N^2-6N+11}{3(N-2)(N-1)} \right) F \\ \frac{(N+1)(N+2)}{3(N-2)(N-1)}F \\ \frac{-(N-3)(N+1)(N+2)}{6(N-2)(N-1)} H \\  \frac{-(N-4)(N+1)}{3(N-2)(N-1)} H \\
\end{array} \right) \ , \ \
V_{A}^{(-)} = \left( \begin{array}{cc} F \\ \frac{N-4}{2(N-2)} F \\  \frac{-1}{N-2} F   \\ \frac{2}{N-2} F \\ -\frac{1}{2} H \\  0 \\ 
\end{array} \right) \ , \ \
V^{(-)}_{SA} = \left( \begin{array}{cc} \frac{(N-3)(N+2)}{4}F \\ \frac{-(N-3)}{N-2}F \\  \frac{-(N-4)}{2(N-2)} F \\ \frac{-(N+2)}{2(N-2)} F   \\ 0 \\ -\frac{1}{2}H \\
\end{array} \right) \ . \ \
\end{align}
The sum rule is consistent with the crossing kernels in the literature \cite{Baume:2021chx}\cite{He:2023ewx}.\footnote{Note that some literature uses the convention which is different from ours in several signs in the conformal block (e.g. in \cite{Chester:2019wfx}).}

Now we can set up the semi-definite problem from the unitarity constraint (e.g. $\Delta \ge \frac{d-2}{2}$ for spin-zero operators)\footnote{To be more precise, we impose the constraint slightly above the unitarity bound on spin-zero operators to speed-up the optimization, which does not affect the final results as far as we have checked.  See \cite{Chester:2019ifh} for a related discussion. \label{footnotegap}} and give the constraint on conformal data. It is important to realize that the single correlator bootstrap presented here can be studied for non-integer $N$ and $d$. Except for the tiny gap in the spin-zero sectors mentioned in footnote \ref{footnotegap}, our conformal bootstrap does not have any artificial gap in any sector: it is a vanilla setup that must be true in any unitary conformal field theories with the specified symmetry. The constraint, therefore, is universal.

\subsection{$O(d)$ bootstrap in $d$ dimensions}
Let us first show the bound of the conformal dimensions of spin-zero operators in $\phi \times \phi$ OPE in $O(d)$ anti-symmetric matrix model in $d$ dimensions with $d=3.98,3.8,3.4$ and finally $3$. The numerical bootstrap was performed in the same setup as our previous paper \cite{Nakayama:2017vdd} but with $\Lambda=17$.\footnote{All the numerical computations in this paper are done on a single 8-core desktop computer. It uses a customized version of cboot \cite{cboot} as an interface with SDPB  \cite{Simmons-Duffin:2015qma}\cite{Landry:2019qug} as a semi-definite problem solver.} For each plot presented below, we have sampled about 20 $\Delta_\phi$ and interpolated the bound by using a Mathematica function.\footnote{The Mathematica interpolation and the scale of the vertical axis may make the kink less visible. The raw data is available upon request.}

It has been observed many times that if we study the bound on the spin-zero operators that are singlet under the symmetry group (we call $S$ sector), the bound coincides with the bound of the singlet spin-zero operator with the enhanced $O(N_V)$ symmetry, where the $O(N_V)$ is the maximal symmetry that external operator scalar operator can possess as a vector representation (see e.g. \cite{Poland:2011ey}\cite{Reehorst:2020phk}\cite{Li:2020tsm}). In our case, the anti-symmetric tensor of $O(N)$ can be embedded in the vector of $O(N(N-1)/2)$ so it coincides with the bound of the $O(N(N-1)/2)$ vector model. We will explicitly see this in the following plots (with black lines).

\begin{figure}[htbp]
	\begin{center}
		\includegraphics[width=12.0cm,clip]
  {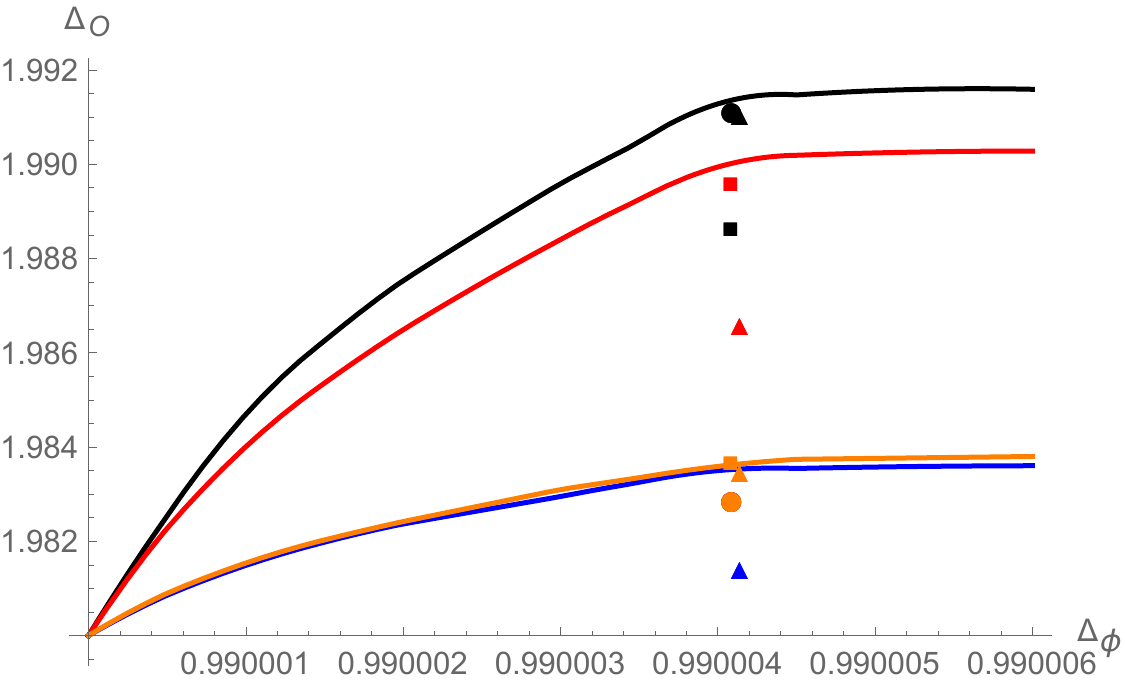}
	\end{center}
	\caption{Conformal bootstrap bound with $O(3.98)$ symmetry in $d=3.98$. Black $S$. Red $A_4$. Blue $T$. Orange $B_4$. The horizontal axis is the conformal dimension of a spin-zero operator in the anti-symmetric representation.
 Predictions of the $\epsilon$ expansions are presented by Circle $V$, Box $M_-$, and Triangle $M_+$.
The kink in the $B_4$ sector seems saturated by $M_-$ at $(0.99000408, 1.98365)$. The kink in the $A_4$ sector could be saturated by $M_-$ at $(0.99000408, 1.98955)$, but it may be accidental. The kink in the $T$ sector is unidentified. 
}
	\label{fig:9898}
\end{figure}

We start with $d=3.98$ presented in Fig \ref{fig:9898}, where the $\epsilon$ expansion should be reliable. All $S$, $T$, $A_4$ and $B_4$ sectors have kinks around $\Delta_{\phi} = 0.99004$. The explicit evaluation of formula \eqref{perturbative} tells us that three nontrivial fixed points have $\Delta_{\phi} =  0.99000409$ at $V$, $0.9900414$ at $M_+$, and $0.99000408$ at $M_-$, so they are coincidentally very close.

Compared with the perturbation theory, besides the $O(d(d-1)/2)$ vector model fixed point $V$ saturated by the kink in the $S$ sector, we clearly see that the unstable fixed point $M_-$ is located at the kink in the $B_4$ sector $(0.99000408, 1.98365)$  and potentially also at the kink in the $A_4$ sector $(0.99000408, 1.98955)$. We later, however, argue that this may be just a coincidence and that the kink in the $A_4$ sector does not correspond to the matrix model fixed point.
Another mystery is the kink in the $T$ sector, where the fixed point in the anti-symmetric matrix model has $\Delta_{T} = 1.98287, 1.98144$ and $1.97974$ at $V$, $M_+$ and $M_-$ respectively, and none of them seems to saturate the bound or even close. 

The situation seems similar to the ``Platonic CFT" discussed in \cite{Stergiou:2018gjj} (see \cite{Kousvos:2018rhl}\cite{Kousvos:2019hgc} for further investigations) when they studied the conformal bootstrap with the cubic symmetry. There, they observed kinks whose origin in the $\epsilon$ expansion is not immediately clear. In our case, one possibility is that in addition to the anti-symmetric matrix $\phi$, the theory has more fields such as $O(N)$ singlet scalar, $O(N)$ vectors, and even fermions. They can give nontrivial fixed points with the same symmetry.

\begin{figure}[htbp]
	\begin{center}
  \includegraphics[width=12.0cm,clip]
  {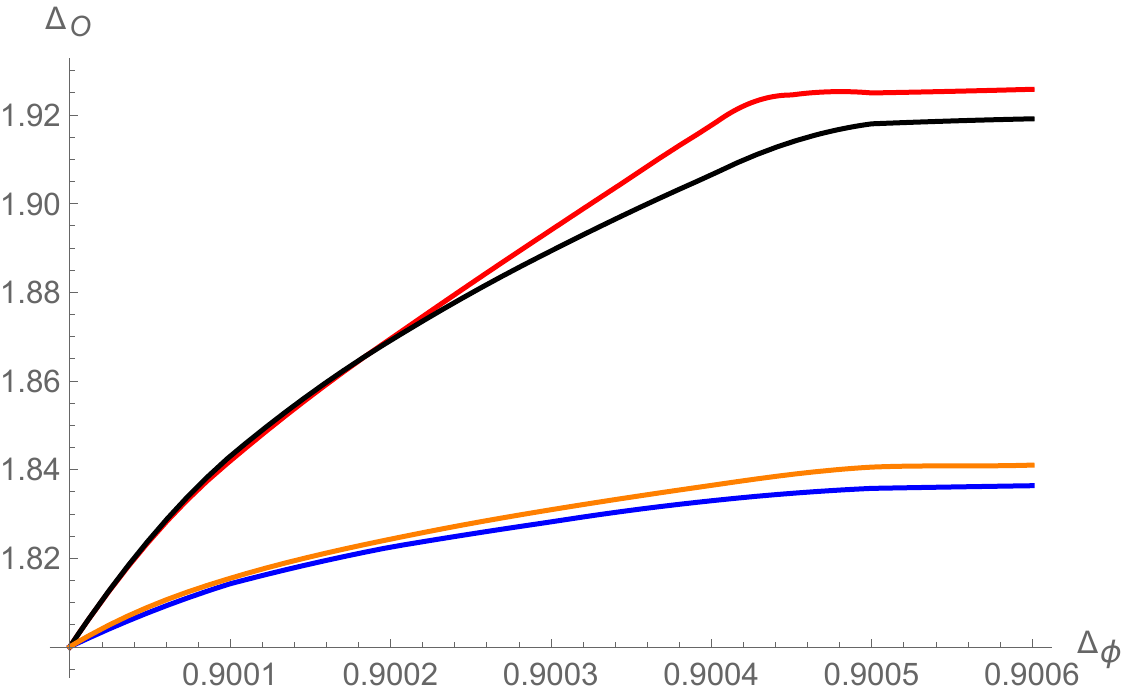}
	\end{center}
	\caption{Conformal bootstrap bound with $O(3.8)$ symmetry in $d=3.8$. Black $S$. Red $A_4$. Blue $T$. Orange $B_4$.}
	\label{fig:3838} 
\end{figure}

\begin{figure}[htbp]
	\begin{center}
  \includegraphics[width=12.0cm,clip]
  {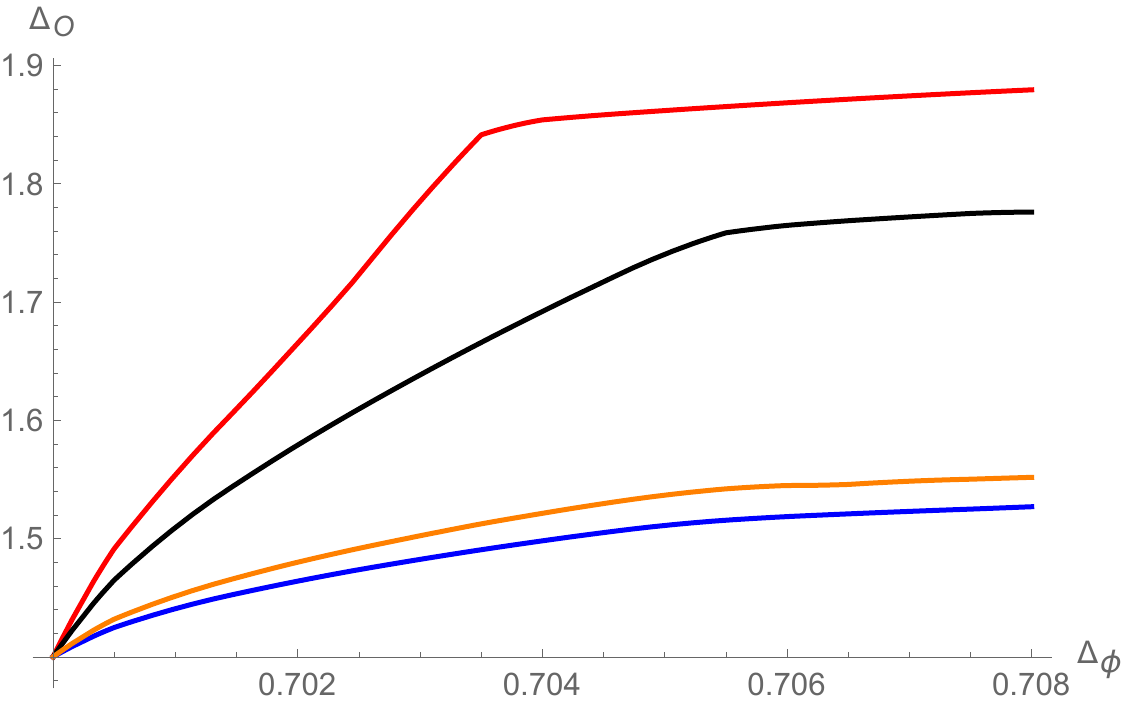}
	\end{center}
	\caption{Conformal bootstrap bound with $O(3.4)$ symmetry in $d=3.4$. Black $S$. Red $A_4$. Blue $T$. Orange $B_4$.}
	\label{fig:3434} 
\end{figure}

\begin{figure}[htbp]
	\begin{center}
		\includegraphics[width=12.0cm,clip]
  {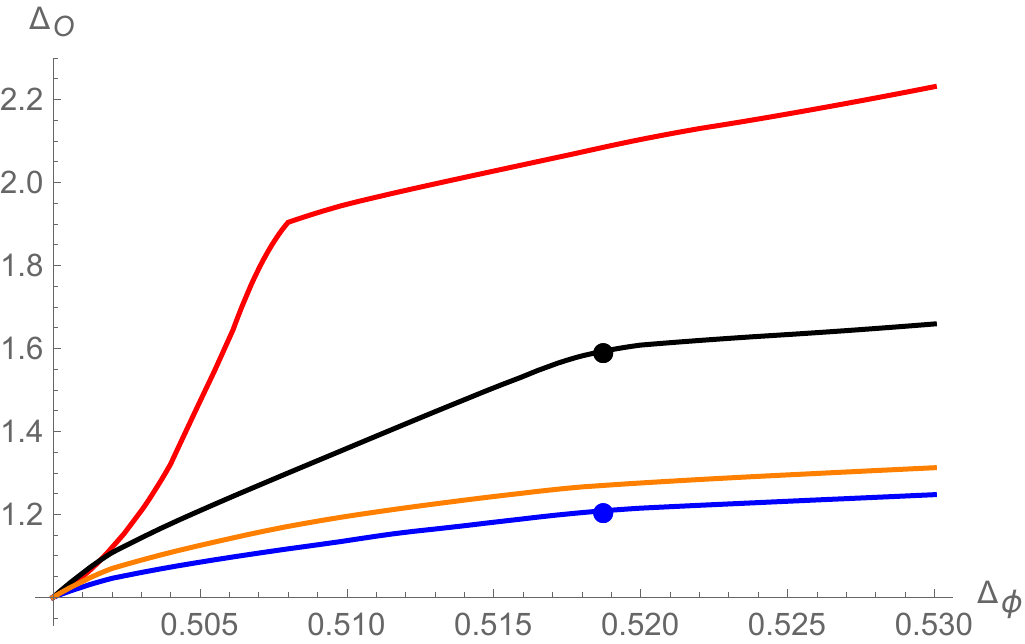}
	\end{center}
	\caption{Conformal bootstrap bound with $O(3)$ symmetry in $d=3$. Black $S$. Red $A_4$. Blue $T$. Orange $B_4$. Circle is $O(3)$ vector model in $d=3$ (Heisenberg fixed point). 
 The kinks in  the $S$ and $T$ sectors are saturated by the $O(3)$ vector model.}
	\label{fig:33}
\end{figure}

If we lower $d$, what we first observe is while the kinks of $S$, $T$, and $B_4$ are located at similar $\Delta_{\phi}$, the location of the kink in the $A_4$ sector tends to deviate toward much smaller $\Delta_{\phi}$.  In the next section, we will see that this rapid change of the location of kink in the $A_4$ sector is caused mainly by the change of $N$ rather than the change of $d$.

Secondly, we observe that the kink in the $B_4$ sector becomes less sharp. This may be understood if we note that one fixed point (e.g. $M_-$ here) cannot saturate two kinks e.g. in the $A_4$ and $B_4$ sectors if the location of $\Delta_\phi$ is different. Nontrivial swapping of the fixed points inside the bound can explain the behavior: now $M_-$ is deep inside the bound and does not saturate it. Or, the fixed point may have disappeared. Note that the (weak) kink in the $B_4$ sector cannot be identified with the vector model fixed point $V$ even though $\Delta_{\phi}$ appears close: such an identification would be inconsistent with the $T$ bound because at the vector model fixed point $\Delta_{T} = \Delta_{B_4}$. 

In the $d \to 3$ limit shown in fig \ref{fig:33}, the bounds in the $S$ and $T$ sectors agree with the bounds in the  $S$ and $T$ sectors of the $O(3)$ vector model studied in \cite{Kos:2014bka}. We naturally expect such behavior because an $O(3)$ anti-symmetric tensor is equivalent to an $O(3)$ vector, and thus the bound of the $O(3)$ anti-symmetric matrix model is the same as the bound of the $O(3)$ vector model (at least for the physical operators; see below on the evanescent operators). 

It, however, makes the kink of the $T$ sector near $d=4$ discussed above, which cannot be the $O(N_V)$ vector model fixed point, even more puzzling. Do we obtain an unidentified conformal fixed point by keeping track of the location of the kink realized by the physical fixed point if we change the dimension $d$ continuously? We do not know a direct answer to this question at this point, but in the next subsection, we will see that the change of $N$ rather than $d$ affects the interpretation of kinks more.

How about the kinks in the $A_4$ and $B_4$ sectors in $d=3$? Since there are no $A_4$ or $B_4$ operators in the $O(3)$ limit of the $O(d)$ symmetry, the physical meaning of the kinks is not immediately obvious. We call them evanescent kinks. Note that we cannot regard the $A_4$ kink as a limit of the $A_4$ operator in the vector model fixed point $V$ simply because $\Delta_{\phi}$ has a different value from the one at the $O(3)$ invariant Heisenberg fixed point. If the kink should be identified with a limit of certain conformal field theories with $O(d)$ symmetry, it must be distinct from the Heisenberg model in any respect.

What we can conclude so far is that it is always safe to study the bound in the $S$ sector and keep track of the symmetry-enhanced vector model fixed point to reach the Heisenberg fixed point. The kinks in the $B_4$ and $T$ sectors may show swapping of fixed points that saturate the bound. The eminent kink in the $A_4$ sector is unidentified. To gain more intuition, in the next subsection, we study conformal bootstrap bound in other $d$ and $N$ to go away from the constraint $N=d$ motivated by the magnetization.

\subsection{More bootstrapping}
The immediate question one may have is what is the meaning of constraining the dimension of operators that do not exist in the $N\to3 $ limit such as $A_4$ and $B_4$. To address this question, let us consider the simpler setup of the $O(N_V)$ vector model in the $N_V \to 1$ limit. The OPE sum rule of the $\langle \phi \phi \phi \phi \rangle $ four-point function, where $\phi$ is a vector of $O(N_V)$) is 
\begin{align}
0 = \sum_{S \in \phi \times \phi} \lambda_S^2 \begin{pmatrix}
 0 \\ F \\ H 
\end{pmatrix}  + 
\sum_{T \in \phi \times \phi} \lambda_T^2 
\begin{pmatrix}
 F \\  (1-\frac{2}{N_V})  \\ -(1+\frac{2}{N_V}) H 
\end{pmatrix}
+\sum_{A \in \phi \times \phi}  \lambda_A^2
\begin{pmatrix}
 -F \\ F \\ - H 
\end{pmatrix}
\end{align} 
and we study the bound on $\Delta_S$ and $\Delta_T$ in the $N\to 1$ limit. 

\begin{figure}[htbp]
	\begin{center}
		\includegraphics[width=12.0cm,clip]
  {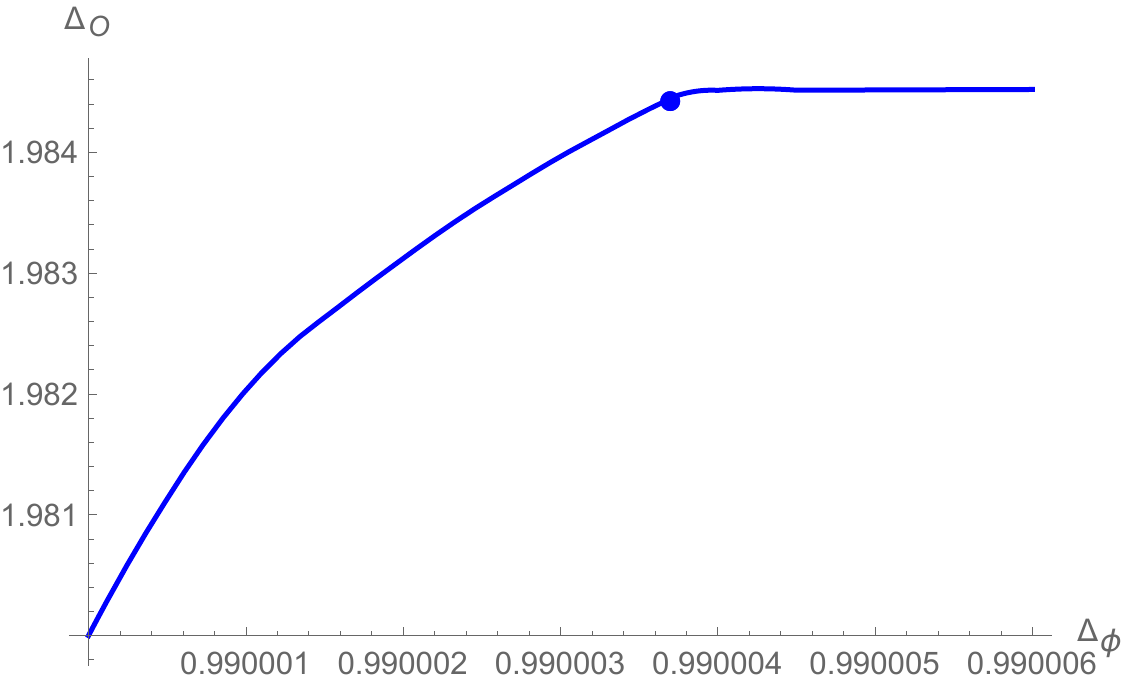}
  \includegraphics[width=12.0cm,clip]
  {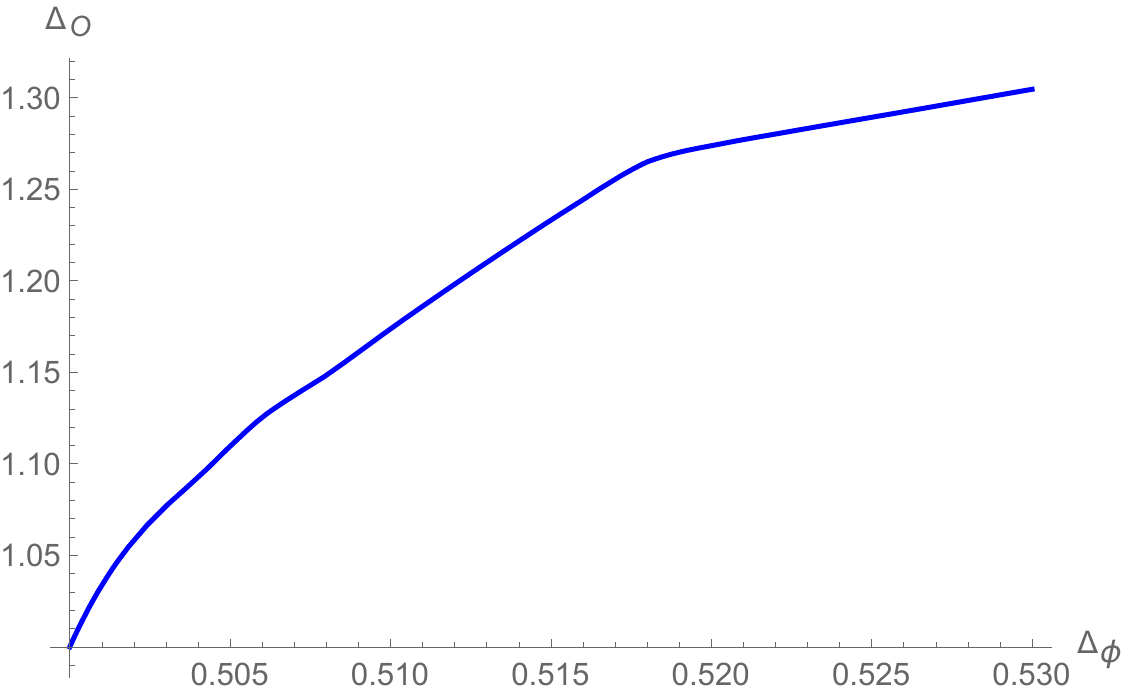}
	\end{center}
	\caption{Conformal bootstrap bound of  ``the $T$ sector with $O(1)$ symmetry" in $d=3.98$ (top) and in $d=3$ (bottom).  The horizontal axis is the conformal dimension of a spin-zero operator in the fundamental representation. We see that kinks are located at the $N_V\to 1$ limit of the $O(N_V)$ vector model fixed point. In $d=3.98$ dimensions, the $\epsilon$ expansion predicts  $(\Delta_\phi,\Delta_T) = (0.9900037,1.98444)$ and shown as circle in the plot.}
	\label{fig:O1T}
\end{figure}

As expected, the bound on the $\Delta_S$ is the same as the bound on the $\mathbb{Z}_2$ case (i.e. Ising case) studied in \cite{ElShowk:2012ht} and we do not report their result here. What seems more nontrivial is the bound on $\Delta_T$. We present the bound in $d=3.98$ and $d=3$ in Fig \ref{fig:O1T}. The kink in $d=3.98$ is located precisely at the prediction of the $\epsilon$ expansion in the $N\to 1$ limit (i.e. $(\Delta_\phi,\Delta_T) = (0.9900037,1.98444)$)
 although the symmetric traceless representation (i.e. $T$) does not exist in the Ising model. The kink in $d=3$ seems to be located at the $N=1$ limit of the $O(N)$ vector models in $d=3$.
\cite{Sirois:2022vth} finds the same conclusion in the multi-correlator conformal bootstrap.

One lesson we can draw here is that bootstrapping the scaling dimensions of evanescent operators (here $T$ sector in the $O(N_V)$ vector model) may give a kink located at the physical fixed point. This observation makes the unidentified kink found in the $A_4$ sector of the $O(3)$ anti-symmetric bootstrap much more interesting because the location of the kink is not the Heisenberg fixed point. Can it indicate another physical fixed point with the $O(3)$ symmetry with an evanescent $A_4$ operator?

Bootstrapping of $O(N)$ anti-symmetric matrix model with $N\neq d$ may have some interest in particular when $N$ is an integer. We show the results of $O(3)$ and $O(6)$ bootstrap in $d=3.98$, and  $O(4)$ and $O(6)$  bootstrap in $d=3$. They may also shed some light on the unidentified kinks in the previous subsection.

\begin{figure}[htbp]
	\begin{center}
  \includegraphics[width=12.0cm,clip]
  {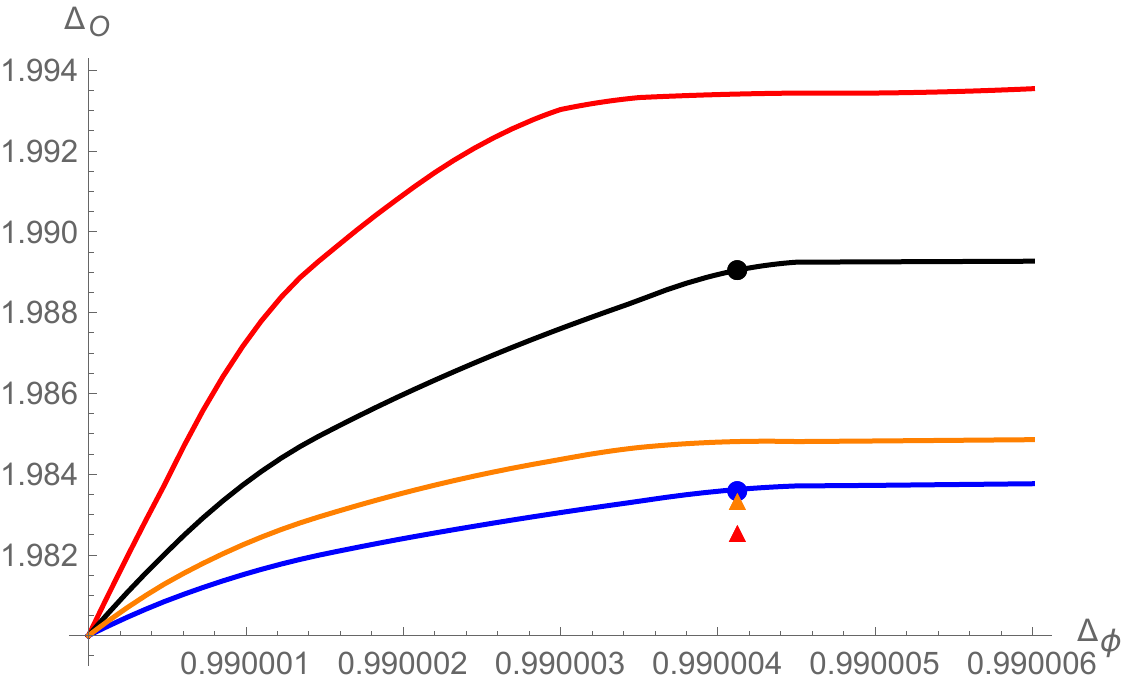}
	\end{center}
	\caption{Conformal bootstrap bound with $O(3)$ symmetry in $d=3.98$. Black $S$. Red $A_4$. Blue $T$. Orange $B_4$.  The horizontal axis is the conformal dimension of a spin-zero operator in the anti-symmetric representation. The bounds in the $T$ and $S$ sectors are the same as those of the vector model and the kinks are located at the $O(3)$ vector model $V$ represented by Circle. We do not know a candidate for the kinks in the $A_4$ and $B_4$ sectors. The triangles are predictions of evanescent operators from $M_+$. The predictions from $M_-$ are not shown here, which are located at $\Delta_{\phi}=0.99$ and violate the bound.}
	\label{fig:98n3} 
\end{figure}

Let us begin with the $O(3)$ case in $d=3.98$. We see that the bounds in the $T$ and $S$ sectors are the same as those in the $O(3)$ vector model as was the case in $d=3$. What is more interesting is we have nontrivial kinks in the $A_4$ and $B_4$ sectors. The location of the kink in the $A_4$ sector is different from the one in the $O(3)$ vector model fixed point, as was the case in $d=3$. We do not know any immediate candidate of the kink, and the origin is of mystery. This indicates that the appearance of the unidentified kink is not associated with the $d=3$ limit, but rather universal in the $N\to 3$ limit.

Some comments on evanescent operators are in order. Unlike in the $N_V\to 1$ limit of the $O(N_V)$ vector model discussed at the beginning of this subsection, the anomalous dimensions of the evanescent operators in perturbation theory are different between $V$ and $M_+$ and between $M_-$ and the Gaussian fixed point. Thus we do not have a definite prediction of the conformal dimensions of the evanescent operators as an unambiguous limit. Moreover, the limit of $M_-$ is inconsistent with the bootstrap bound because while $\gamma_{\phi}$ tends to zero, anomalous dimensions of $A_4$ and $B_4$ remain positive.

\begin{figure}[htbp]
	\begin{center}
  \includegraphics[width=12.0cm,clip]
  {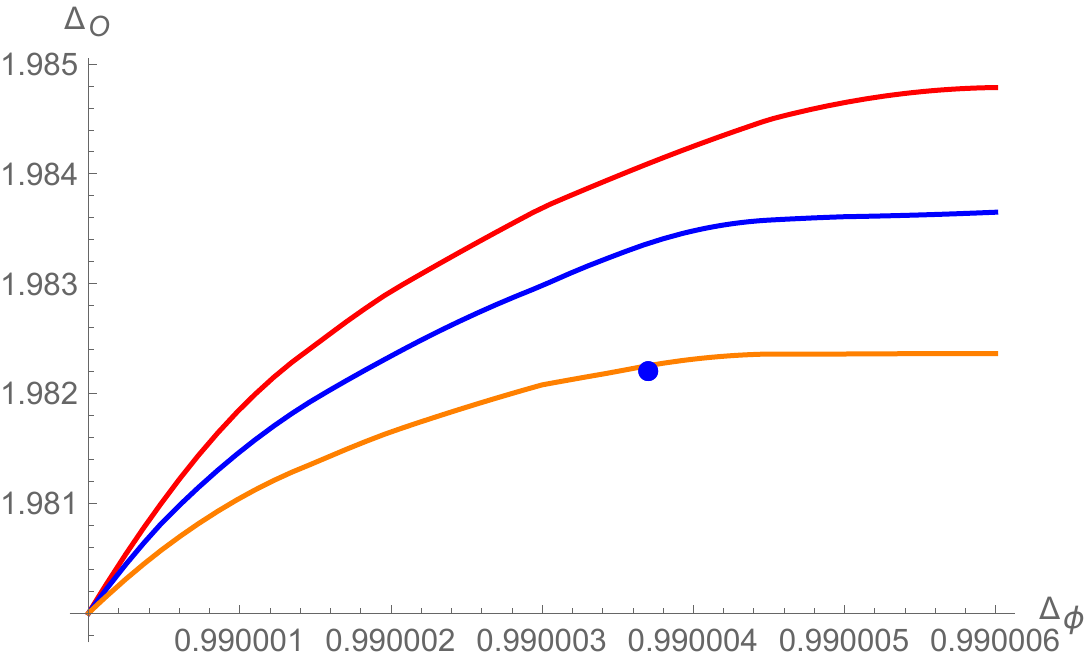}
	\end{center}
	\caption{Conformal bootstrap bound with $O(5)$ symmetry in $d=3.98$.  Red $A_4$. Blue $T$. Orange $B_4$. Circle $O(15)$ vector model. The bound in the $S$ sector is the same as the one in the $O(10)$ vector model, so it is omitted. The kink in the $B_4$ sector seems saturated by the $T$ operator of the $O(10)$ vector model. Note $M_{\pm}$ are complex fixed points and not shown in the plot.}
	\label{fig:98n5} 
\end{figure}

\begin{figure}[htbp]
	\begin{center}
  \includegraphics[width=12.0cm,clip]
  {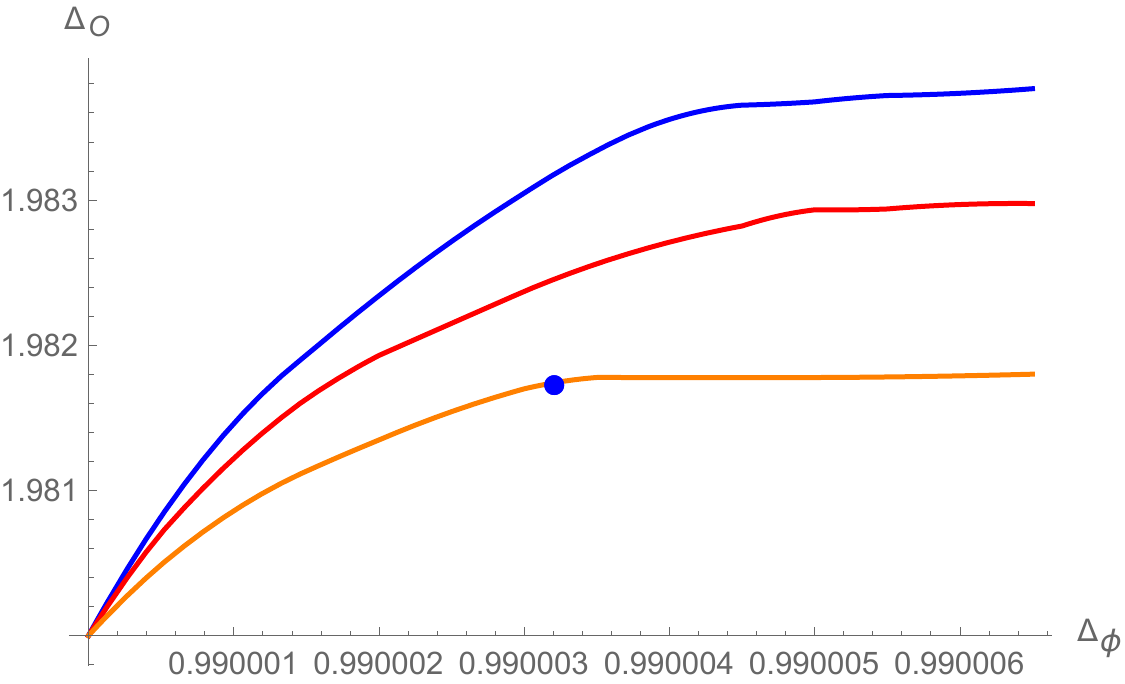}
	\end{center}
	\caption{Conformal bootstrap bound with $O(6)$ symmetry in $d=3.98$.  Red $A_4$. Blue $T$. Orange $B_4$. Circle $O(15)$ vector model. The bound in the $S$ sector is the same as the one in the $O(15)$ vector model, so it is omitted. The kink in the $B_4$ sector seems saturated by the $T$ operator of the $O(15)$ vector model. Note $M_{\pm}$ are complex fixed points and not shown in the plot.}
	\label{fig:98n6} 
\end{figure}

Next, let us discuss the $O(5)$ and $O(6)$ cases in $d=3.98$ presented in fig \ref{fig:98n5} and fig \ref{fig:98n6}. The $\epsilon$ expansion 
 of the anti-symmetric matrix model shows that the matrix model fixed points $M_{\pm}$ are complex fixed points and the only real fixed point is the $O(15)$ vector model.
  We observe that the kink in the $B_4$ sector is precisely saturated by the $T$ operator of the $O(10)$ and $O(15)$ vector model. It is uncommon to observe an enhancement of the symmetry in the bound of the non-singlet operator. The fact that the fixed point that saturates the $B_4$ sector is switched from $M_-$ (when $N\sim 4$) to $V$ (when $N =5, 6$) should coincide with $5> N_c$ when $\epsilon$ is small.
  
  {In the bound of the $A_4$ sector, we see a kink at a larger value of $\Delta_\phi$, which is not predicted by the real fixed points in the $O(6)$ anti-symmetric matrix model. Together with the results at $N=3$ and $N=3.98$, we conjecture that the kink in the $A_4$ sector is not directly related to the anti-symmetric matrix model.} We also see a kink in the $T$  sector. {As in $N = 3.98$ (but unlike in $N=3$ which is saturated by the vector model fixed point), a potential kink in the $T$ sector is not saturated by any known fixed point. We do not know its origin.}

\begin{figure}[htbp]
	\begin{center}
  \includegraphics[width=12.0cm,clip]
  {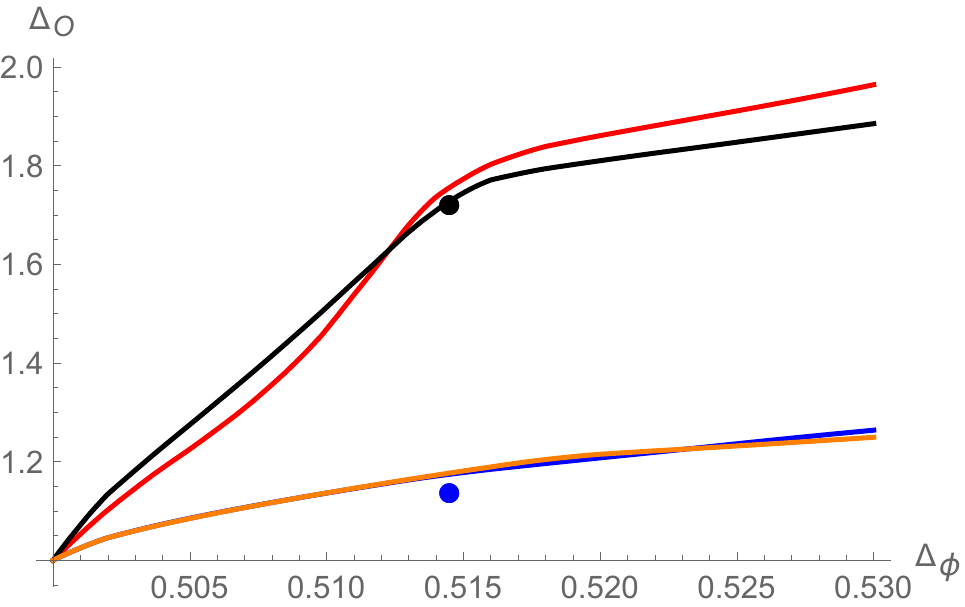}
	\end{center}
	\caption{Conformal bootstrap bound with $O(4)$  symmetry in $d=3$. Black $S$. Red $A_4$. Blue $T$. Orange $B_4$.  Circle $O(6)$ vector model. While the $O(6)$ vector model is located at the kink in the bound of the $S$ sector, it does not saturate the bounds in the $T$ and $B_4$ sectors.}
	\label{fig:3n4}
\end{figure}

\begin{figure}[htbp]
	\begin{center}
  \includegraphics[width=12.0cm,clip]
  {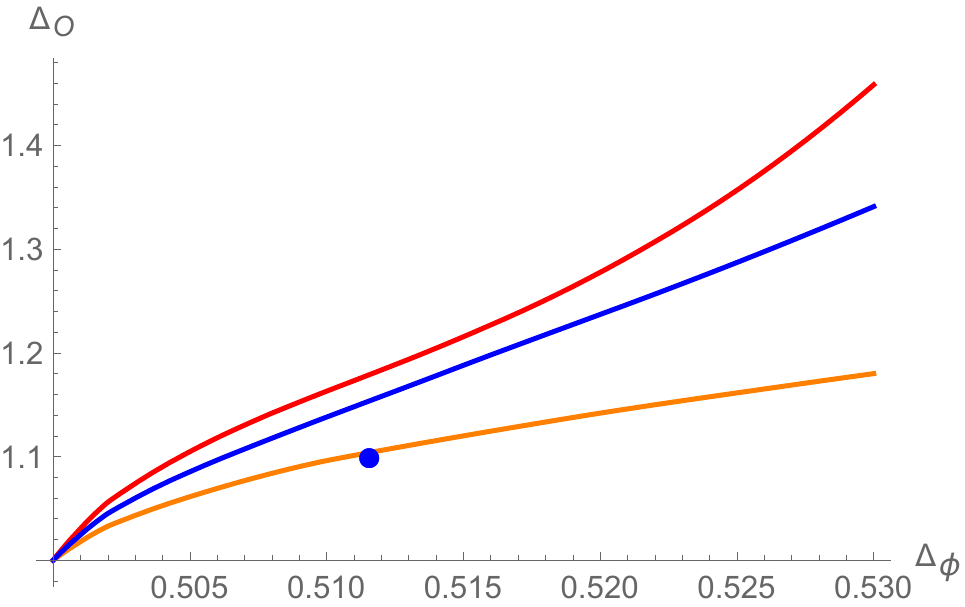}
	\end{center}
	\caption{Conformal bootstrap bound with $O(5)$ symmetry in $d=3$. Red $A_4$. Blue $T$. Orange $B_4$. Circle $O(10)$ vector model. The bound in the $S$ sector is the same as the vector model, so it is omitted. The $O(10)$ vector model is located closely at the kink in the $B_4$ sector. The kink in the $A_4$ sector located at $\Delta_{\phi} = 0.542$ and the weak kink in the $T$ sector located around $\Delta_{\phi} = 0.55$ are not shown in the plot.}
	\label{fig:3n5}
\end{figure}

\begin{figure}[htbp]
	\begin{center}
  \includegraphics[width=12.0cm,clip]
  {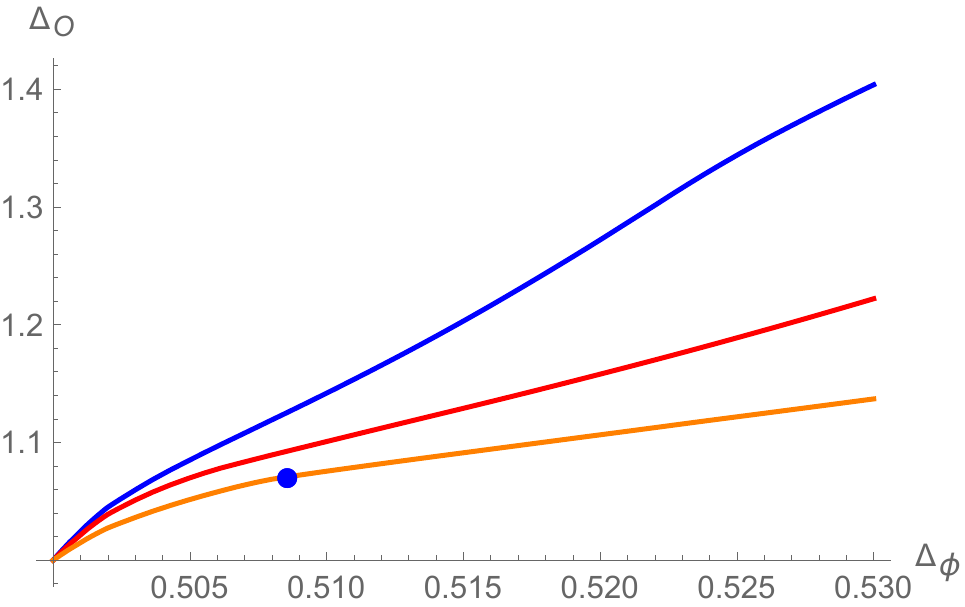}
	\end{center}
	\caption{Conformal bootstrap bound with $O(6)$ symmetry in $d=3$. Red $A_4$. Blue $T$. Orange $B_4$. Circle $O(15)$ vector model. The bound in the $S$ sector is the same as the vector model, so it is omitted. The $O(15)$ vector model is located at the kink in the $B_4$ sector. The kink in the $A_4$ sector is located at $\Delta_{\phi} = 0.58$ and is not shown in the plot.}
	\label{fig:3n6}
\end{figure}

We finally present physically interesting bound with $O(4)$, $O(5)$ and $O(6)$ symmetries in $d=3$. See Fig \ref{fig:3n4}, \ref{fig:3n5} and \ref{fig:3n6}. They could be realized in nature as the Landau-Ginzburg model with $O(N)$ anti-symmetric matrix order parameter. In the $O(4)$ case, we find that the $T$ operator in the vector model fixed point with the enhanced $O(6)$ symmetry does not saturate any bound. {The kink in the $A_4$ sector around $\Delta_\phi = 0.516$  and another one in the $B_4$ sector around $\Delta_\phi = 0.519$ may indicate a nontrivial matrix model fixed point such as $M_-$ or potentially $M_+$.} Compare them with the $O(3.98)\sim O(4)$ in $d=3.98$ in Fig \ref{fig:9898}, where $M_-$ seems to be located at kinks of $A_4$ and $B_4$. There, we have also seen that the $T$ operator in the vector model fixed point does not saturate any bound.

In contrast, if we look at the $O(6)$ case in $d=3$ presented in Fig \ref{fig:3n6}, we see that the $T$ operator in the $O(15)$ vector model fixed point is located at the kink in the bound of the $B_4$ sector. In addition, we find a weak kink in the $T$ sector around $\Delta_{\phi} = 0.525$.  The potential switching of the saturation of the $B_4$ kink to the $O(15)$ vector model fixed point may suggest $N_c< 6$ in $d=3$. While it is not shown in Fig \ref{fig:3n6}, there exists a weak kink in the $A_4$ sector around $\Delta_{\phi} = 0.58$. Currently, we do not know the origin of the kinks in $T$ and $A_4$ sectors.

{The analogous study of the $O(5)$ case in $d=3$ suggests that the weak kink in the $B_4$ sector is located closely but not as precisely at the $O(10)$ vector model (see Fig. \ref{fig:3n5}). It appears slightly below the bound, which may indicate that $N_c \sim 5$ in $d=3$. 
We also observe a weak kink around $\Delta_\phi = 0.542(2)$ in the $A_4$ sector and another weak kink around $\Delta_{\phi}= 0.55$ in the $T$ sector, whose origins are not immediately identified.}

\begin{table}[!ht]
    \centering
     \scalebox{1.0}{
    \begin{tabular}{|l|l|l|l|l|l|l|l|l|}
    \hline
        $d$  &$3$ & $3.98$ & $3$ & $3.98$ & $3$ & $3.98$ & $3$ & $3.98$ \\ \hline
          $N$    &$3$ & $3$ & $4$ & $3.98$ & $5$ &$5$ & $6$ & $6$ \\ \hline\hline 
        $S$ & $V$ & $V$ & $V$ & $V$ & $V$ & $V$& $V$ & $V$ \\ \hline
        $T$ & $V$ & $V$ & very weak  & ? & very weak & ? & very weak & ? \\ \hline
        $A_4$ & ? & ? & ? & $M-$? & weak & ? & weak & ? \\ \hline
        $B_4$ & ? & ? & ? & $M_-$($M_+$?) & $V$? &  $V$ &  $V$ & $V$ \\ \hline
    \end{tabular}}
    \caption{(Conservative) Identification of kinks with known fixed points.}
	\label{table:kink}
\end{table}

We have summarized a conservative identification of kinks with known 
 fixed points in table \ref{table:kink}. Since we do not know the conformal dimensions in $d=3$ other than the vector model fixed point (such as those in $M_{\pm}$) precisely, we do not present any conjectural identification in this table.

The table suggests that the switching of the fixed point saturating the kink is more sensitive to $N$ rather than $d$.
If we accept the hypothesis that the identification of the kink shares a similar feature with $N$ rather than $d$, we predict that the unstable fixed point $M_-$ in $O(4)$ anti-symmetric matrix model is located at $\Delta_{\phi} = 0.519(1)$.


\section{Discussions}

In this paper, we have studied the conformal bootstrap constraint in the $O(d)$ anti-symmetric matrix model in $d$ dimensions, varying $d$ as a continuous parameter. We have also bootstrapped $O(4), O(5), O(6)$  anti-symmetric matrix model in $d=3$, aiming at physical applications. We have found many kinks, some of which are identified with the known fixed points, and some of which are yet to be identified. 

To address the original question of what is the best way to approach the Heisenberg model in $d=3$ dimensions from the anti-symmetric matrix model, we find that keeping track of the vector model fixed point with the enhanced $O(d(d-1)/2)$ symmetry is always safe. We do not, however, find a good way to keep track of the most stable fixed point $M_+$ within the single correlator vanilla bootstrap studied in this paper.

While we have focused on the $O(d)$ anti-symmetric matrix model in $d$ dimensions, it has become more evident that the vast landscape of the $O(N)$ anti-symmetric matrix model in $d$ dimension when $d\ne N$ is yet to be explored. Some sample studies in section 3 reveal that one kink associated with a known fixed point saturating one bootstrap bound becomes not saturating when we change $N$ or $d$. Is it a continuous process or a discontinuous process? Is it related to the collision of the fixed points in the renormalization group flow? 

One obvious future direction is to use the more recent techniques in numerical conformal bootstrap such as ``navigator functions" \cite{Reehorst:2021ykw} and ``skydiving" \cite{Liu:2023elz} to explore the mixed correlation functions to isolate ``islands" in anti-symmetric matrix models. In contrast to the vanilla setup used in this paper, the gap assumption necessary for the analysis is nontrivial and potentially requires more physical input.

Another limitation of our study is that we have studied the bound in smaller $\Delta_\phi$. It is possible if we further increase $\Delta_\phi$, we may find a kink (or even the second or higher ones \cite{Nakayama:2019jvm}\cite{He:2020azu}\cite{Ghosh:2023onl}) near or above $\Delta_{S} \sim d$ \cite{Nakayama:2017vdd}. We may expect they are realized by non-Landau-Ginzburg theories such as gauge theories or fermionic theories, but such interpretations are, if any, still conjectural.

Understanding the stability of the conformal field theory realized at the boundary of the conformal bootstrap bound, most importantly at kinks or islands, should be further studied. For instance, in our problem, we naturally expect that if we identify a certain kink with the unstable fixed point (e.g.  $M_-$), the spectrum read from the extremal functional technique should include two relevant singlet operators. In reality, it is not necessarily so. The absence of the second relevant singlet operator in the extremal functional was observed in various examples (e.g. in the models studied in \cite{Nakayama:2014lva}). It is important to understand why the crossing equation is not so sensitive to the (non)existence of the second relevant operates in the spectrum. 

We have restricted our analysis in $d\ge 3$, but the study of the fixed point in $O(N)$ ant-symmetric matrix model in $d<3$ (or even $d>4$) may be of interest. For instance, in a recent paper \cite{Jepsen:2020czw} they found much more surprising features in the renormalization group flow by including $\phi^6$ interactions. Bootstrapping these theories seems of great interest although the unitarity or even the conformal invariance of these fixed points may be questioned.

In this paper, we have studied the consequence of treating the magnetization as a two-form rather than a vector of the ``internal" $O(d)$ symmetry. It, however, becomes much more nontrivial if we further demand that the magnetization is a space vector (or two-form). In particular, if we introduce the so-called dipolar interaction through the exchange of the dynamical magnetic field, we are forced to regard the magnetization as a two-form under the spatial rotation. 

In \cite{FisherAharony}\cite{AharonyFisher}\cite{AF-PRBII} (see also \cite{Gimenez-Grau:2023lpz}\cite{Nakayama:2023wrx} for more recent discussions), the magnetization was treated as a space vector in $4-\epsilon$ dimensions with the dipolar interaction. In this approach, the dipolar interaction becomes effectively local, leading to the transverse constraint $\partial_i \phi^i = 0$. On the other hand, if we treat the magnetic field as a two-form since we cannot dualize the two-form in $4-\epsilon$ dimensions in a meaningful sense, the interaction remains non-local. This will make the problem more nontrivial than what we encounter in this paper.

Finally, we wonder if the anti-symmetric matrix model with an integer $N>3$ may have some physical applications. Typically, the matrix index is tied up with the space index, and if so it is unphysical to consider $N>3$. The more abstract vector space may be called for.
See \cite{Antonov:2013aka}\cite{Antonov:2017pqv} and reference therein.\footnote{Let us, nevertheless, add one daydream here. If we allow the analytic continuation, we can think of four-dimensional Lorentz indices. Suppose our fundamental theory is a membrane theory in three dimensions with an anti-symmetric matrix on it. The conformal invariance (but why?) then would indicate that the symmetry must be $O(4)$ (rather than $O(11)$!). Would this explain four-dimensional space-time?}

\section*{Acknowledgements}
This work is in part supported by JSPS KAKENHI Grant Number 21K03581.

\end{document}